\documentclass{aa}
\usepackage{boxedeps,longtable,dcolumn,times}
\def\plumi{\raisebox{0.4ex}{$\scriptscriptstyle{\pm}$}}
\def\et{{et\thinspace al.}\ }                     
\def\kms{km\thinspace s$^{-1}$ }                  
\def\kmx{km\thinspace s$^{-1}$}                   
\def\onerule{\noalign{\medskip\hrule\medskip}}        

\begin{document}

\thesaurus{03        
              (04.03.1) 
              (11.03.1) 
              (11.03.4)
              (11.07.1)
              (11.06.1)}

\title{Velocity structure of the dwarf galaxy population in the
Centaurus cluster}

\thanks{Based on observations made at the European Southern Observatory,
La Silla, Chile.}

\author{P.~Stein\inst{1} 
\and    H.~Jerjen\inst{2}
\and    M.~Federspiel\inst{3}}

\institute{
     Departament d'Astronomia i Meteorologia, Universitat de Barcelona,
     Avenida Diagonal 647, E--08028 Barcelona, Spain.
     E-mail: paul\@@pcess2.am.ub.es
\and Mt. Stromlo and Siding Spring Observatories, Australian National 
University, Private Bag, Weston Creek PO, ACT 2611, Canberra,
     Australia. E-mail: jerjen\@@mso.anu.edu.au
\and Astronomisches Institut der Universit\"at Basel, 
     Venusstrasse 7, CH--4102 Binningen, Switzerland. E-mail: 
     federspielm\@@ubaclu.unibas.ch}

\offprints{M.~Federspiel (Universit\"at Basel)}


\maketitle

\begin{abstract} 
Based on the photometric survey of the inner region of the Centaurus
cluster (Jerjen \& Dressler 1997a) we measured redshifts for a deep,
surface brightness limited sample of galaxies using the MEFOS
multifibre spectrograph at the ESO 3.6m telescope. With the new data
set radial velocities for 120 centrally located cluster members become
available which is equivalent to 78\% of all known cluster galaxies in
the region brighter than $B_{\rm T}$=18.5. The relevant aspect of this
investigation is that new redshifts for 32 dwarf galaxies have been
measured, rising the total number to 48. We investigate the prominent
bimodal velocity distribution of Centaurus in more detail, discussing
the very different characteristics of the velocity distributions for
the main Hubble types E\&S0, spirals, Im\&BCD, and dE\&dS0. The
nucleated, bright dwarf ellipticals are the only galaxies with a
Gaussian-like distribution centred at 3148$\plumi$98\,\kms. The
remarkable coincidence of this velocity with the mean velocity of Cen30
and the redshift of NGC$\,$4696 in particular strongly suggests a
connection of the dE\&dS0s to the gravitational centre of the Centaurus
cluster and/or to the cluster dominant E galaxy. The application of
statistical tests reveals the existence of a population dwarf galaxies
bound to NGC$\,$4696.  The dynamical parameters for the two velocity
components suggest that Cen30 is the real Centaurus cluster whereas
Cen45 can only be a loosely bound group of galaxies. This conclusion is
followed up with a type-mixture analysis. All results are fully
consistent with the cluster-group scenario. Whether Cen45 is merging
with the cluster or is located in the close background remains
unclear. We show that the poorness of Cen45 represents an intrinsic
problem which makes it difficult to approach this question.

\end{abstract}
   
\section{Introduction}

Lucey et al.~(1986a, hereafter LCD) discovered a remarkable bimodal
velocity distribution for the classical Hubble types E, S0, and spirals
in the Centaurus cluster. The two velocity components were found to be
roughly centred at 3000 \kms and 4500 \kms and were denoted as Cen30
and Cen45, respectively. Two fundamentally different scenarios have
been suggested to explain the phenomenon. (1) Centaurus may be just a
superposition of two spatially separated clusters at relative distances 
to Virgo of 1.67 and 2.38, respectively (Lynden-Bell et al.~1988, Faber 
et al.~1989). If true, large peculiar velocities for the two clusters
would be the consequence in good agreement with the large-scale streaming 
motion of nearby galaxies towards the "Great Attractor" (Dressler et 
al.~1987, Lynden-Bell et al.~1988), a large concentration of mass $\sim$ 
5$\cdot 10^{16}$\,M$_{\sun}$ with its gravitational potential well close 
to the galaxy cluster Abell 3627, as was discovered recently by Kraan-Korteweg
et al.~(1996).  
(2) A less spectacular but equally attractive explanation for the
bimodality is that we are dealing with a young, dynamically unrelaxed
galaxy cluster consisting of two unequal subclumps (Lucey \& Carter
1988, hereafter LC). The less massive component Cen45 is about to merge
with the main cluster component Cen30 and the large velocity difference
of $\sim$ 1500\,\kms for the components is interpreted as due to the
gravitational infall of Cen45 into the major cluster centre.
Although some new results have been published on the Centaurus cluster
recently (Dressler 1993; Jerjen 1995; Jerjen \& Dressler 1997b), a
conclusive explanation for the origin of the bimodal velocity
distribution could not yet be given. Centaurus is the closest galaxy
cluster to the Local Group showing this kind of feature and thus the
best candidate to explore such a phenomenon. A clarification of the
situation is highly desirable for a better understanding of structure
formation in the Universe on the scale of galaxy clusters as well as on
larger scales such as the proposed ``Great Attractor'' supercluster.
 
Moreover, there is another challenge offered by the Centaurus
cluster. From kinematical studies of the clusters in Virgo (Binggeli et
al.~1993) and Fornax (Held \& Mould 1994) we know that the velocity
distribution of early-type dwarfs might differ significantly from that
of the giants. This could be due to cluster formation effects and/or
evolution of dwarf galaxies in high density environments.  In
particular, the observed clumpiness of low density regions in galaxy
clusters and the marked spatial and velocity segregation of galaxy
types (e.g. Stein 1997) suggest the occurrence of significant late
infall of spiral galaxies from the cluster outskirts onto the core
(Binggeli \et 1987). These infalling late-type galaxies are likely to
be initially surrounded by a bound population of dwarf satellites, as
was found to be the case for the field (Vader \& Sandage 1991). It
seems to be clear that a significant fraction of bound (dwarf) galaxies
can be found in clusters (Ferguson 1992), but it is still not known
whether all reach the central cluster region as companions of spirals
or whether there are other evolutionary effects involved.  It has been
proposed to extract dynamical information about a cluster using dwarf
galaxies as test particles in the cluster potential well (Binggeli \et 1987),
given that their gravitational pull on neighbouring giant
galaxies is negligible.  In the case of relaxed clusters with a central
dominant galaxy up to 5\% of the cluster members were found to be
possibly bound to the central system (Gebhardt \& Beers 1991;
Merrifield \& Kent 1991; but see Blakeslee \& Tonry 1992). So far the
only studies about bound populations involving distinctions between
dwarf and giant galaxies have been Ferguson's (1992) and Binggeli's
(1993) analysis of the Virgo cluster.
 
The aim of the present study is to explore the velocity distributions
of different galaxy types in the Centaurus cluster with special weight
on the dwarf families. This shall help to give us more insight into
the kinematic properties of this complex galaxy aggregation and of 
type-dependent cluster dynamics in general.
 
The sample selection is discussed in Sect.~2. A technical description
of the spectroscopic observations and of the data reduction follows in
Sect. 3. The resulting new redshifts were then used to check the
validity of previous membership assignments based on morphological
information alone (Sect. 4). We present our analysis of the data in
Sections 5 and 6, where type-dependent velocity distributions are
interpreted and the existence of bound galaxies is investigated. Main
results are summarized in Sect. 7.

\section{Sample selection}
With the deep photometric Centaurus cluster survey (Jerjen 1995) an
extensive list of new cluster dwarf galaxies became available including
accurate positions and high morphological resolution. These latter
quantities are based on a fine-scale 20-inch Las Campanas du Pont plate
covering the central 1.5$\times$1.5 degrees of the cluster. The
probability of cluster membership was worked out based on pure
morphological criteria and is given as 50\% or higher for each galaxy
included in the catalogue. For our redshift project we selected all
galaxies of the outcoming Centaurus cluster catalogue (Jerjen \&
Dressler 1997a, hereafter CCC) with a mean effective $B$-band surface
brightness {\it SB}$_{\rm eff}$ brighter than
25.0\,mag\,arcsec$^{-2}$. This implies our sample contains a
substantial number of newly discovered dwarf galaxies as well as
several mainly giant galaxies with known redshifts (Dickens et
al.~1986, hereafter DCL; LC; Stein 1996). The latter subsample was used
for comparison.

The restriction of our redshift survey onto the inner region of the
Centaurus cluster was not only dictated by the area of the photometric
survey but is also required to overcome uncontrollable cluster sample
contamination by unrelated galaxies located at a Hubble distance of
$\sim$ 4500\,\kmx. In fact, this is a crucial problem for the Centaurus
cluster as already emphasized elsewhere (Lucey et al.~1986b).
 
\section{Observations and data reduction}
Observations were carried out between May 23--26, 1995 at the 3.6~m ESO
telescope at La Silla. We employed the multifibre instrument MEFOS,
which has a circular field of view of 1 degree and 29 fibre arms. Each
arm carries one image fibre of 36$\times$36 arcsec and two spectral fibres
of 2.5 arcsec aperture for simultaneous object and sky acquisition. The
image fibre is meant for the interactive repositioning of spectral
fibres onto the object of choice, prior to the spectral exposure.
For a detailed study about the performance of MEFOS consult Felenbok
\et 1997.

\begin{figure*}[tb]
\SetRokickiEPSFSpecial
\HideDisplacementBoxes
\centerline{\EPSFxsize=17.6cm \EPSFbox{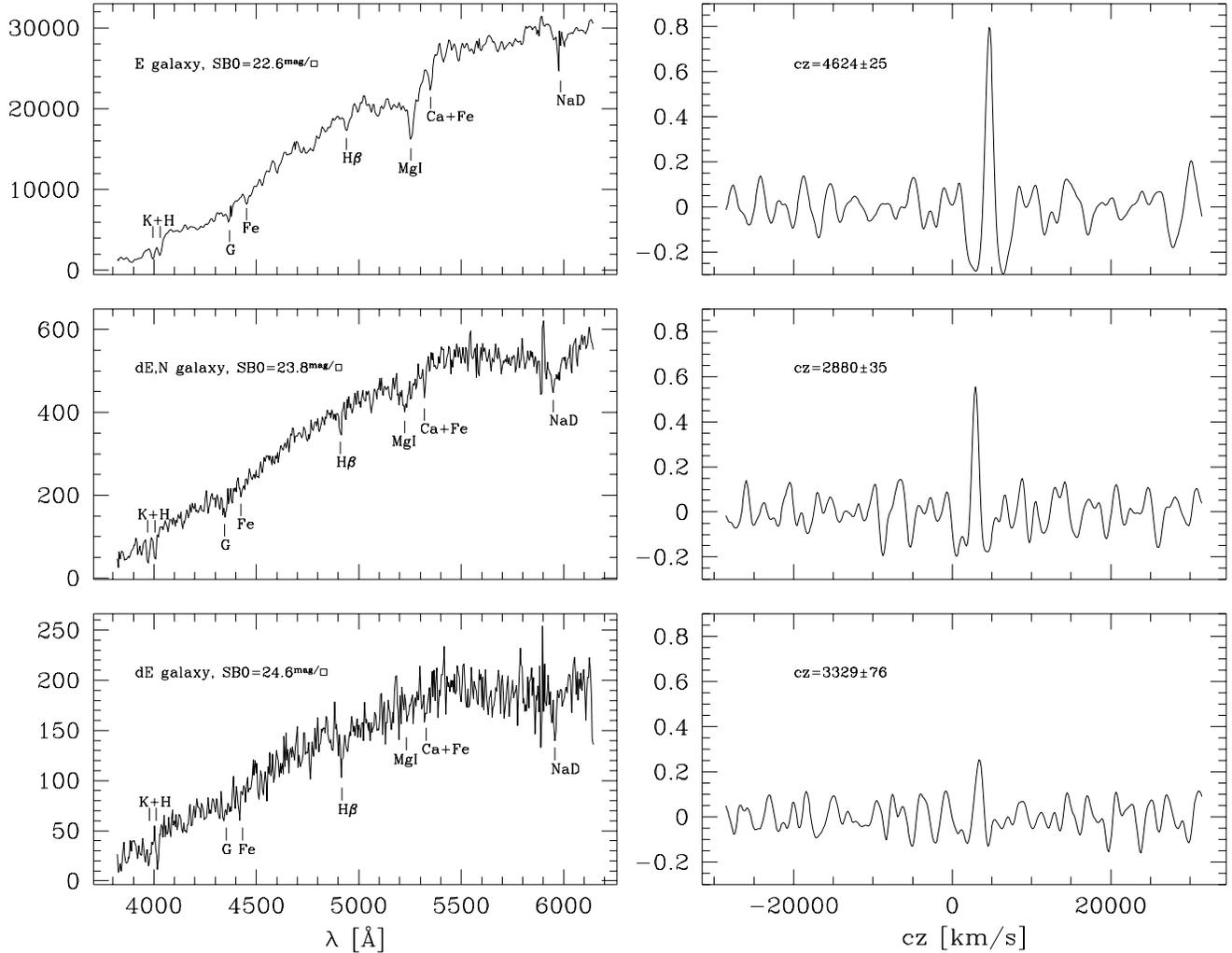}}
\caption{Typical spectra and their cross-correlation functions. The
uppermost spectrum belongs to NGC$\,$4709, the dominant galaxy of
Cen45. Note the good symmetry in the correlation function around its
peak. At the bottom a very low-surface brightness galaxy is shown, for
which a redshift measurement was at the limit of feasibility. The
inferred positions of some of the main absorption lines are marked.}
\end{figure*}     

No beam-switching technique has been applied for subtraction of the
night sky spectral contribution (Cuby \& Mignoli 1994), because the
traditional technique with fibre transmission correction (given by the
signal under the O{\small I} emission line at 5577.4\,\AA) gave
suitable results. The 512 pixels of the CCD covered the wavelength
range 3800 to 6100\,\AA.  A total exposure time between 5400 and
7200\,s per field was chosen.

Reduction of the spectroscopic data was done using the MIDAS package on
an IBM AIX/RS6000.  All frames were carefully corrected for
pixel-to-pixel variations in sensitivity (flat-fielded).  Then an
optimal extraction algorithm was applied (Horne 1986), which was able
to correct for most of the cosmic ray hits. No correction for stray
light nor cross-talk between fibres (Lissandrini \et 1994) was
applied, because it could have influenced the precision of the optimal
extraction. The resulting one-dimensional spectra were then calibrated
in wavelength and the positions of night-sky emission lines were checked
against their expected wavelength. Note that no appreciable systematic
deviation in the position of night-sky lines was found, as opposed to
the findings of Felenbok \et 1997, who comment on their problems with
observations taken roughly one year before us.

Prior to sky subtraction the signal of each spectrum (including sky
spectra) had to be scaled with respect to the intrinsic transmission 
efficiency of the corresponding fibre. To determine the transmission
efficiency of each fibre the signal under the 5577.4 \AA\ night-sky 
line was used, taking an average for each fibre over all observed fields. 
Creation of a mean sky spectrum was done taking all available sky
spectra from the corresponding pair of exposures.

\subsection{Emission-line redshifts}

First, those galaxies were selected which had at least two clearly
visible emission lines, mostly O{\small II}, H$\beta$ and O{\small
III}. A Gaussian superposed onto a quadratic polynomial was then used
to fit each line interactively. The final emission line redshift was
computed as the unweighted mean over all $N$ emission lines present in
one galaxy spectrum. Errors in each measurement were taken to be 100
\kmx, independently of line strength, because the main factor of
uncertainty involved had been found to be the wavelength calibration
(Stein 1996). Thus, the errors of the emission line redshifts are taken
as 100$/\sqrt{N}$.

\subsection{Cross-correlation redshifts}

Prior to cross-correlation redshift measurements, residual features
around the positions of strong night-sky lines had to be removed, as
well as all galaxy emission lines. Next, a procedure of spectra
preparation was followed which was virtually the same as described in
Tonry \& Davis (1979). See Stein (1996) for details about template
selection. Typical spectra and peaks of the obtained cross-correlation
functions are shown in Fig.~1.

Table 1 lists the efficiency (number of redshifts obtained/number of
observed galaxies) of the instrument together with the reduction
procedure up to a given limiting effective surface brightness, as well
as the corresponding limiting apparent total magnitude for point-like
sources. The latter quantity was estimated taking into consideration
the fibre radius of 1.25 arcsec, and under the assumption of an
exponential surface brightness profile, a mean scale length of 2.9
arcsec, a mean effective radius of 3.6 arcsec (Jerjen \& Dressler
1997a) and for the case that the fibre had been ideally centred onto
the galaxy. Two fields were excluded from the analysis because of their
much lower yield due to problems with the autoguider.

The limit in surface brightness which could be reached here confirms
that projects aiming at the systematic measurement of redshifts for 
dwarf galaxies in nearby clusters are feasible, even with a
multi-fibre instrument. The only requirement is a large field of view. 

\tabcolsep0.5cm
\begin{table}[t]
\begin{tabular}[c]{rrr}
\onerule
\multicolumn{1}{c}{{\it SB}$_{\rm eff}$} & 
\multicolumn{1}{c}{$B_{\rm T}$} & 
\multicolumn{1}{c}{efficiency}\\
\multicolumn{1}{c}{(1)} & 
\multicolumn{1}{c}{(2)} & 
\multicolumn{1}{c}{(3)} \\
\onerule                    
      $<$ 23.0 &       $<$ 20.4 &     100 \%\\
      23.0 -- 23.5 & 20.4 -- 20.9 &     100 \%\\
      23.5 -- 24.0 & 20.9 -- 21.4 &      84 \%\\
      24.0 -- 24.5 & 21.4 -- 21.9 &      73 \%\\
      24.5 -- 25.0 & 21.9 -- 22.4 &      26 \%\\
\onerule
\end{tabular}      
\caption{Efficiency in redshift determination for galaxies of different
effective surface brightnesses. For comparison purposes, the
hypothetical $B$ magnitude of a point-like source which on the average
would feed the fibres with the same amount of light is also
given. Optimal positioning of the fibres is hereby assumed.}
\end{table}    

\begin{table*}[bt]
\caption{Morphological membership assignment compared to redshift data
(explanation see text)}
\begin{tabular}{rr|rrr|rrr}
\multicolumn{2}{c|}{}&\multicolumn{3}{c|}{Im\&BCD}&\multicolumn{3}{c}{dE\&dS0}\\
\multicolumn{2}{c|}{membership}& m & b & membership & m & b &membership\\
\multicolumn{2}{c|}{(morph)}   &   &   & (redshift) &   &   & (redshift)\\
\multicolumn{2}{c|}{(1)}       &(2)&(3)&(4)         &(5)&(6)&(7)\\\hline

        1=&100\%     & 7 & 1 & 88\%       &14& 2 &       88\%\\
        2=&75\%      & 1 & 0 & 100\%      & 3& 0 &       100\%\\
        3=&50\%      & 3 & 5 & 38\%       & 4& 3 &       57\%\\\hline
\end{tabular}
\end{table*}

\subsection{Scaling and checking the data}
Correction for the earth motion was then carried out, leading to 
heliocentric corrected redshifts for both the cross-correlation and
the emission-line redshifts. 

For 32 galaxies the redshifts obtained here could be compared to those
obtained previously by one of us (Stein 1996). Note that
cross-correlation errors for the brightest galaxies in Stein (1996)
might be slight underestimations, due to the fact that the template had
been constructed using a sample of these same bright galaxies. Thus,
the scaling factor between internal (Tonry \& Davis 1979) and external
errors is determined using only galaxies in the calibration sample with
errors of 20 \kms or above. With an error scaling factor of 1.7 the
differences between redshifts in both datasets are perfectly consistent
with the resulting external errors. A zero-point shift of 26 \kms was
then applied to the cross-correlation data.

Only for six galaxies it was possible to obtain both cross-correlation
and emission-line redshifts.  Again, the distribution of differences
is in good agreement with the expectations, given the uncertainties in
both measurements.  Finally, a weighted mean of emission and absorption
(cross-correlation) redshifts was taken.

\section{Membership: morphology versus redshift}
We measured redshifts for 115 galaxies. The data are listed in
Tables\,5a, 5b, and 5c combined with data from the CCC such as
morphological type, total apparent $B$\,magnitude, and
{\it SB}$_{\rm eff}$. Galaxies with no previous redshift measurement are
subdivided into two lists according to cluster members (Table\,5a) and
background objects (Table\,5b). Table\,5c contains the data for galaxies
which already had a redshift measured and are confirmed cluster members.

Among our galaxies there are 101 with velocities smaller than 5414
\kms. This velocity corresponds to the 3$\sigma$ upper limit for the
velocity distribution of Cen45 (LCD) and was used as a cut-off to
discriminate between cluster members and background galaxies. Actually,
the remaining 14 velocities lie between 10\,000 and 38\,000 \kms
which makes a separation unambiguous. Our cluster sample was
complemented by 19 redshifts taken from the literature (DCL; LC; Stein
1996). Finally, the dataset upon which the following analysis is based
consists of redshifts for 120 cluster members located in the central
cluster area. A weighted average was taken in case of multiple
redshifts, after homogenization of zero-point-shifts and scaling of
errors among the four sources.  The completeness of this data set is
very high with respect to apparent magnitude and effective surface
brightness. Redshifts are available for 96 percent of all known cluster
galaxies brighter than $B_{\rm T}$=17.5. At the limit of $B_{\rm
T}$=18.5 our redshift sample is still complete to 78 percent. This
latter magnitude corresponds to $M_{B_{\rm T}}=-$15.3, assuming a
cluster distance modulus of $(m-M)_{\rm Cen}$=33.79 (Jerjen \& Dressler
1997b) and including an extinction term of $A_{\rm B}$=0.42.  For {\it
SB}$_{\rm eff}$ the completeness levels are 93 percent at
23.5$\,B$\,arcsec$^{-2}$ and 78 percent at 24.5\,$B$\,arcsec$^{-2}$,
respectively.

\begin{figure*}[th]
\SetRokickiEPSFSpecial
\HideDisplacementBoxes
\ForceWidth{13.5cm}
\BoxedEPSF{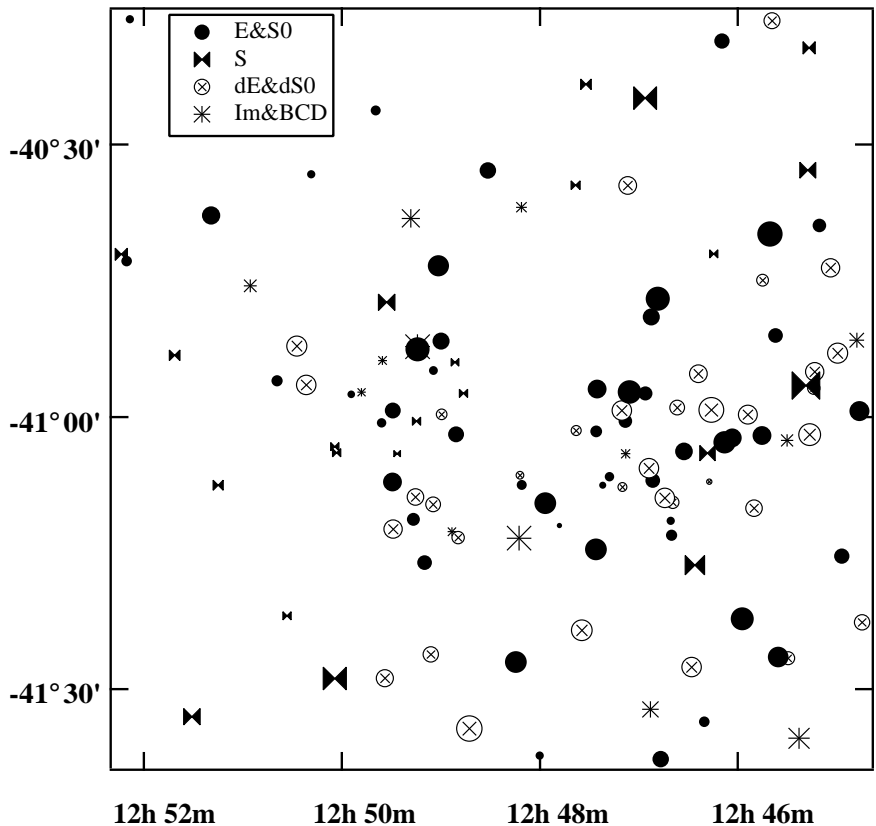}
\hfill \parbox[b]{4cm}{
\caption{ A map of all Centaurus cluster member galaxies with known
redshifts. The symbol size is inversely proportional to the redshift.}}
\end{figure*}      

Tables\,5a and 5b contain data for 50 galaxies with new measured
velocities among which 36 galaxies are cluster members according to our
selection criterion. Their Hubble type mixture is
\mbox{(E\&S0/S/Im\&BCD/dE\&dS0) =} (1/3/11/21).  The composition of the
background sample is (0/3/6/5). For the latter sample the morphological
information may not necessarily be true any more because the
classification had been done under the assumption of cluster
membership. The new measured redshifts shall be used to estimate the
accuracy of the morphological based cluster membership for dwarf
galaxies in the CCC. For this purpose we compare in Table\,2 the number
of cluster members and background objects for the two dwarf families
individually. In column\,1 we give the three membership classes as
listed in the CCC. Columns\,2 and 3 give the observed numbers of
late-type dwarfs of a particular membership class divided into members
(m) and background (b) according to their redshifts. From columns\,2
and 3 we derive the fraction of real cluster members in column\,4 which
can be compared to the percentages listed in column\,1. The same
analysis is done for the early-type dwarfs in the columns 5--7. As can
be seen, the numbers in columns\,1, 4 and 7 are in good agreement. The
100\% case is slightly overestimated (what one naturally would expect)
whereas the 75\% case is clearly underestimated. Combining class 1 and
2, the membership assignment has a success rate of 90\% for both dwarf
families. Overall, these results strongly support the idea that
morphological criteria are an excellent tool to discriminate between
cluster and background galaxies even for clusters at the distance of
Centaurus if one works with high resolution material and spatially well
isolated clusters.

\section{Velocity distributions}

An illustration of the 3D distribution of the Centaurus galaxies is
given in Fig.~2 using the two coordinates and radial velocity. In
projection, Centaurus shows a pronounced bimodal substructure along the
East-West direction. At RA$\sim$12$^{\rm h}$48$^{\rm m}$30$^{\rm s}$ a
vertical void in the galaxy distribution separates the two cluster
subclumps. The significance of this so-called Centaurus Gap is shown
based on Lee statistics and Monte Carlo cluster models (Jerjen
1997). The large Western clump is mostly populated with early-type
galaxies and dominated by the brightest cluster member, the E/S0 galaxy
NGC$\,$4696 at a velocity 2961$\plumi$26 \kms (weighted mean of three
sources). The late-type galaxies can be found preferentially in the
smaller Eastern concentration which leads to a morphological
segregation in the 2D distribution for giants (LCD) and dwarfs
described in detail by Jerjen (1995). But it is obvious that the
morphological mixture also varies as a function of redshift. The low
velocity range (large symbols in the plot) is governed by early-type
galaxies. Going to higher redshifts (small symbols) the mixture changes
in favour of spirals and Im\&BCDs.

\begin{figure}[ht]
\SetRokickiEPSFSpecial
\HideDisplacementBoxes
\ForceWidth{8cm}
\BoxedEPSF{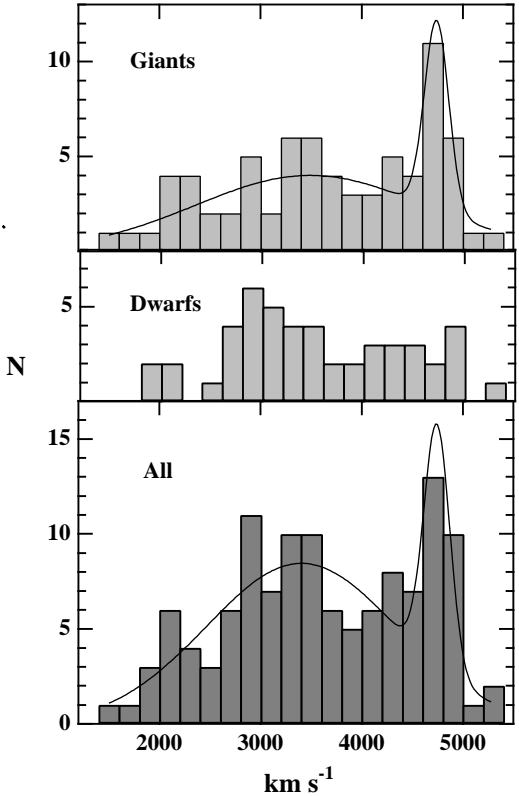}
\caption{The velocity distributions for the giant, the dwarf and the
combined sample. Solid lines represent the best fitting double
Gaussian functions.}
\end{figure}     
 
This morphological segregation with redshift becomes clearer when
focusing onto the radial velocity information only.  Fig.~3 shows the
velocity distributions for giants, dwarfs, and for the complete sample
binned into 200\,\kms intervals. Dwarfs and giants both exhibit
irregular shapes which appear to be totally incompatible.  An
explanation for these puzzling distributions will be found thanks to
the better morphological resolution we have for our sample (see Fig.~4
below).  Only the giant galaxy sample shows a weak indication for a
bimodality which is highly amplified by the dwarfs in the combined
sample. To quantify the bimodal structure of the giant and complete
samples we fitted a double Gaussian function to the data. The total
number of sample galaxies was kept fixed and the errors were assumed to
be $\sqrt{N}$. The best-fitting parameters are listed in Table\,3. Both
parameter sets are in good agreement but are slightly different from
the estimates of LCD.  Their cluster parameters have been included for
comparison. They were derived from a $<1^\circ$ sample of giant
galaxies complete down to {\em G}$_{26.5}$~=~16.5.  Definitely, the
parameter which discriminates best between the two velocity components
is their richness, i.e. the number of galaxies assigned to each
component at our completeness level in mean surface brightness. The
ratio $n_{30}/n_{45}$=4.7 gives a clear hint to the different nature of
Cen30 and Cen45. Cen30 is much richer in galaxies and appears as a
typical galaxy cluster when its large velocity dispersion of
933$\plumi$118\,\kms is compared to the velocity dispersion
distribution for Abell clusters (Girardi et al.~1993). On the other
hand, the first conclusion to be drawn from the very low value for
$\sigma_{45}$ is that Cen45 can only be a loosely bound system. A
galaxy group seems to be a much better picture for this velocity
component rather than a cluster. We like to emphasize that these
results are based on pure statistical and dynamical arguments and
therefore independent from the underlying spatial structure of the
Centaurus cluster.

\tabcolsep0.2cm
\begin{table}[tbp]
        \caption{Best parameters for the double Gaussian fit}
        \begin{tabular}{rrrrr}
        \hline
&Giants          &Early           &All          &Giants \\
&                &types           &             &(LCD)\\
&\multicolumn{3}{c}{($r<0.75^\circ$)}           &($r<1^\circ$)\\
\hline
$n_{30}$     &57$\pm$5    &  62$\pm$10  &99$\pm$7     & \\
$\mu_{30}$   &3491$\pm$225&3170$\pm$174 &3397$\pm$139& 3130\\
$\sigma_{30}$&1135$\pm$200& 738$\pm$175 &933$\pm$118 & 780\\
$n_{45}$     &15$\pm$5    & 11.5$\pm$10 &21$\pm$7     & \\
$\mu_{45}$   &4741$\pm$45 &4788$\pm$154 &4746$\pm$43  & 4634\\
$\sigma_{45}$&120$\pm$41  & 345$\pm$144 &131$\pm$43   & 260\\
$\chi^2$     &7.8          &5.4           &10.1          &\\\hline
\end{tabular}
\end{table}

Our first conclusion about the nature of the galaxy aggregations Cen30
and Cen45 predicts a difference in the galaxy type-mixture taking into
account the morphology-density relation for giants (Dressler 1980) and
dwarfs (Binggeli et al.~1987; Ferguson \& Sandage 1988; Binggeli et
al.~1990). A higher fraction of early-type galaxies is expected in
Cen30 than in Cen45 which shall be examined. Due to the overlapping
Gaussian profiles it is not possible to associate all individual
galaxies to one particular component. However, we can employ the best
fitting pair of Gaussian profiles from the complete galaxy sample to
determine the probability of each galaxy to be member of one particular
component.  Tables\,5a and 5c contain the column prob$_{45}$ which
gives the probability of being a Cen45 member. A value of 0.00 implies
a 100\% probability of being a Cen30 member. Therefrom we calculate
statistically the type-mixtures which are shown in Table\,4.  As
expected, Cen30 is much richer in early-type galaxies than Cen45 with a
74\% portion as compared to 48\%.  In order to understand the prominent
difference in the early to late-type dwarf ratio (dE\&dS0/Im\&BCD)
which is 3.6 for Cen30 and 0.5 for Cen45, respectively, one has to
remember that the brightest dE\&dS0s can be found only in clusters or
as close companions to massive parent galaxies. In fact, all known
dE\&dS0s brighter than about $M_{B_{\rm T}}=-$16 are cluster
members. For instance, the brightest dE\&dS0s in Virgo, Fornax, or
Centaurus are $-$18 (Sandage et al.~1985), $-$17.5 (Ferguson \& Sandage
1988), and $-$18 (Jerjen \& Tammann 1997), respectively. On the other
hand, NGC$\,$205 is the brightest dwarf elliptical in the LG with only
$M_{B_{\rm T}}$=$-$15.6 and the nearby Cen$\,$A group has no dwarf
brighter than $M_{B_{\rm T}}=-$13 (Jerjen et al.~1997). Our
completeness limit is about $M_{B_{\rm T}}\sim-$15.5. Therefore, the
dE\&dS0/Im\&BCD ratio is naturally expected to be much larger for a
cluster than for a group population at this luminosity limit.

\begin{figure}[ht]
\SetRokickiEPSFSpecial
\HideDisplacementBoxes
\ForceWidth{8cm}
\BoxedEPSF{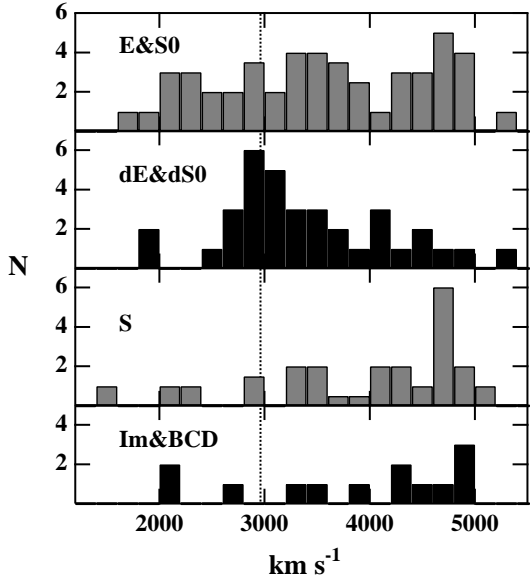}
\caption{The velocity distributions of the Centaurus cluster binned
into different Hubble types. The dotted line indicates the redshift of
the brightest cluster galaxy NGC$\,$4696}
\end{figure}     
 
Another significant difference between groups and clusters is their
dwarf-to-giant ratio (Ferguson \& Sandage 1991). In our case we get
\mbox{(dE\&dS0/E\&S0) =} 0.81 and \mbox{(Im\&BCD/Sp) =} 0.57 for Cen30
and \mbox{(dE\&dS0/E\&S0) =} 0.24 and \mbox{(Im\&BCD/Sp) =} 0.52 for
Cen45, respectively. In terms of environmental differences these numbers
are comparable to the Fornax cluster and the Leo group at our luminosity
limit $M_{B_{\rm T}}=-$15.5 (cf.~Ferguson \& Sandage 1991).  

\begin{table}[tbp]
        \caption{Type mixtures for the two cluster velocity components}
        \begin{tabular}{r@{\hspace{1mm}}r@{\hspace{1mm}}r@{\hspace{1mm}}r@{\hspace{1mm}}r}
        \hline&
\multicolumn{2}{c}{Cen30}&
\multicolumn{2}{c}{Cen45}\\
\multicolumn{1}{c}{Type}&
\multicolumn{1}{c}{number}&
\multicolumn{1}{c}{\%}&
\multicolumn{1}{c}{number}&
\multicolumn{1}{c}{\%}\\
\multicolumn{1}{c}{(1)}&
\multicolumn{1}{c}{(2)}&
\multicolumn{1}{c}{(3)}&
\multicolumn{1}{c}{(4)}&
\multicolumn{1}{c}{(5)}\\
        \hline
        E      &14.17&14&2.32&11\\
        S0     &26.31&27&5.69&27\\
        Spirals&16.21&16&7.29&34\\
        Im\&BCD& 9.21& 9&3.79&18\\
        dE\&dS0&33.10&33&1.90&10\\\hline
        All    &99.00&100&21.00&100\\
        \hline
\end{tabular}
\end{table}

In Fig.~4 the velocity histograms for different Hubble types are shown.
Intermediate types as e.g.~S0/Sa were divided with half weight to both
types. E\&S0 galaxies populate the range 1700-5300\,\kms smoothly
with no obvious centre. In contrast, the dE\&dS0s which are mostly
nucleated dwarf ellipticals are highly concentrated. A least-squares
fit of a Gaussian profile gives $\mu_{dE\&dS0}$=3148$\plumi$98\,\kms
with a velocity dispersion of $\sigma_{dE\&dS0}$=530$\plumi$88
\kmx. There is a remarkable coincidence between the derived mean
velocity and the central velocity of the Cen30 cluster component and in
particular with the velocity of NGC$\,$4696. This strongly suggests a
dynamical link between these bright early-type dwarfs and the cluster
potential and/or the dominant galaxy in Cen30. We will investigate this
issue in Section\,6 in more detail.  It has been claimed that the
dE\&dS0s galaxy class is the most strongly clustered of all (e.g.~Vader
\& Sandage 1991).  Our result from the third dimension (velocity) is a
full confirmation of this 2D-based conclusion.

Like the E\&S0s, the cluster spirals are also distributed over the a
large velocity range 1400-5200\,\kmx.  However, in this case there is
a slight preference to velocities higher than 4000\,\kms (median
value $4350$\,\kms with a peak at 4700\,\kmx.  Below the median
velocity each bin contains an average galaxy number of 0.8. Above this
velocity the value is 2.5 times higher. The velocity distribution of
our late-type dwarf sample compares very well with that of the
spirals. Their distribution is flat covering the range 2000-5000\,\kms
without any obvious concentration. The median velocity of the whole
sample is 4300\,\kmx.

\section{Bound companions}
A remarkable feature of Fig.~4 is that most early type dwarfs are
distributed with a nearly Gaussian-like shape around a velocity of
about 3000\,\kms with a significantly smaller velocity dispersion than
the cluster itself. This rises the question of whether there is a
population of dynamically bound satellites around NGC\,4696. Note that
while the general presence of bound companion galaxies in clusters has
been convincingly proven (Ferguson 1992), there is still uncertainty
about whether these satellites are mainly bound to central dominant
galaxies or not (Gebhardt \& Beers 1991; Merrifield \& Kent 1991, and
references therein).

\begin{figure}[hb]
\SetRokickiEPSFSpecial
\HideDisplacementBoxes
\ForceWidth{8cm}
\BoxedEPSF{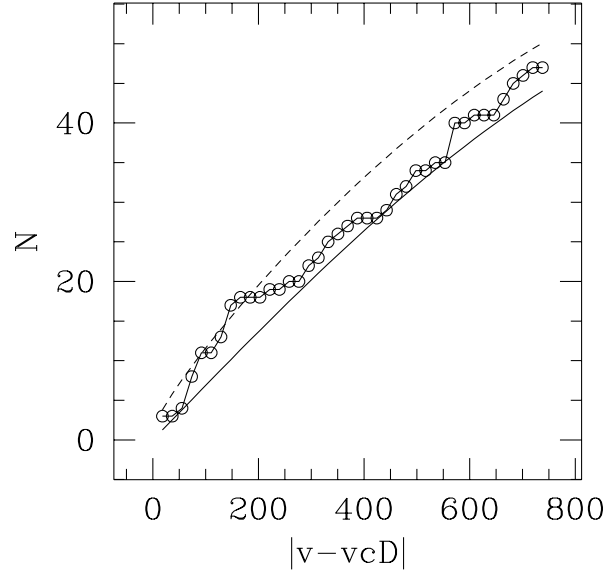}
\caption{``Indicator'' Test for the presence of bound galaxies around
NGC\,4696. The number of galaxies found below a given velocity
difference (circles), resp. expected in
the mean (continuous line) and at most (95 \% level, dashed line) is
plotted against the velocity difference relative to NGC\,4696. Only
early type galaxies with cz $<$ 4250\,\kms have been included 
in the analysis.}
\end{figure}     

To assess the significance of the overabundance of low-velocity
galaxies around NGC\,4696, we applied a slightly modified version of
the ``Indicator'' Test (Gebhardt \& Beers 1991). This test compares the
number of galaxies in a range of velocities around the central galaxy
with the number expected if they were drawn from the overall cluster
(Gaussian) velocity distribution.

Because of the bimodal velocity distribution of Centaurus, and given
that late types are not expected to be normally distributed, we
selected those 65 galaxies of types E, S0, dE, dS0 with velocities less
than 4250\,\kms as pertaining to Cen30.  We then specified the main
population of Cen30 as being drawn from a Gaussian with parameters
taken from Table\,2, column 3, and computed the ``Indicator'' statistic,
taking into account the incompleteness above 4250\,\kmx. See Fig.~5 for
a graphical representation of the ``Indicator'' Test.

The test confirms that those galaxies with radial velocities differing
by less than about 200\,\kms from NGC\,4696 have a 2\,$\sigma$
probability of not being drawn from the overall cluster velocity
distribution. This value is reasonable if it refers to a sample of
bound galaxies, because it lies below what has been measured for the
stellar velocity dispersion of NGC\,4696 itself, i.e. 256\,\kms
(Carollo \et 1993). It can further be seen that for a velocity
difference smaller than 200\,\kms the number of galaxies found is in
excess by about 6 with respect to the Gaussian hypothesis. Considering
that our sample contains of the order of 100 galaxies in Cen30, there
is a good agreement with Merrifield \& Kent's (1991) findings that 5 \%
of the galaxies are bound.

\begin{figure}[ht]
\SetRokickiEPSFSpecial
\HideDisplacementBoxes
\ForceWidth{8cm}
\BoxedEPSF{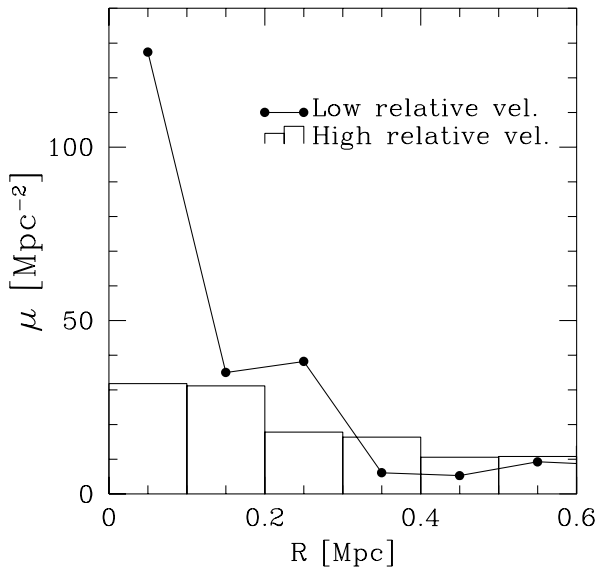}
\caption{Rough surface density profiles for sample galaxies with
$|$v$_{NGC\,4696}$-v$| > 250$\,\kms (histogram) and $< 250$\,\kms
(line), respectively.  Projected distance R of each galaxy to NGC\,4696
has been calculated assuming a Hubble constant H$_0$ =
50\,\kmx\,Mpc$^{-1}$.  Densities have been normalized to the same
number of galaxies in both samples.}
\end{figure}     

Note, however, that the results from the ``Indicator'' statistic are
only true under the assumption that the moments of a Gaussian are good
estimates for the cluster velocity distribution. In particular, this
test is sensitive to the value of the dispersion for the underlying
Gaussian, which is affected by relatively large uncertainties. For this
reason we compared also mean projected separations from NGC\,4696 of
candidate bound companion galaxies with other cluster members. Given
that the sample of bound companion candidates has been selected based
on redshift information only, this procedure gives an independent,
qualitative check of the assumption.  The reader might judge about the
degree of concentration of the two samples towards the central dominant
galaxy with the help of Fig.~6, where rough surface density profiles
are shown for both samples. A correction for incompleteness below
RA\,=\,12$^{\rm h}$~44$^{\rm m}$ has been applied, and the densities
have been normalized to the number of galaxies in the richer sample. In
fact, the two histograms diverge appreciably for R $<$ 100 kpc,
i.e. galaxies with lower relative velocity are preferably located
closer to the dominant galaxy.

\section{Summary and conclusions}
The present redshift survey focused on the central region of the
Centaurus galaxy cluster. Our aim was to investigate the nature of the
reported bimodal structure in its velocity distribution. For this
purpose we observed redshifts for a highly complete and deep sample of
Centaurus cluster galaxies including a substantial number of
dwarfs. Since dwarf galaxies show a more extreme behaviour with respect
to the environment than giants, they offer a powerful diagnostic for
the underlying conditions.  We found significant differences in the
velocity distributions of individual Hubble-types. The dwarf
ellipticals are the only galaxies which are strongly clustered in
velocity with a Gaussian-like distribution at a mean redshift
of 3148$\pm$98\,\kms comparable to the centre of Cen30. This indicates that
this galaxy type is carrying information about the mass distribution of
a galaxy compound.  Our results support the conclusions of earlier
works (e.g.~Vader \& Sandage 1991) that dwarf ellipticals are the
most clustered galaxy type.

We analysed the prominent bimodal velocity distribution of the complete
sample. Dynamical properties of the two velocity components give
evidence that the galaxies associated to Cen30 represent a real galaxy
cluster at $v\sim$ 3400\,\kms whereas Cen45 must be a loosely bound
system. Predictions based on the morphology-density relation for
galaxies offered a test for this conclusion. We carried out a
statistical analysis of the type-mixture of Cen30 and Cen45 galaxies
which confirms the very different nature of the two aggregates. The
true cluster is Cen30 with a large velocity dispersion of
933$\plumi$118\,\kms and a high fraction of early-type galaxies. On the
other hand, Cen45 is a late-type galaxy dominated system with a very
low velocity dispersion of only 131$\plumi$43\,\kms comparable to a poor
galaxy group. With respect to their dwarf-to-giant ratios the velocity
components are as different as the Fornax cluster and the Leo
group. The issue about the spatial location of the two components is
not in reach of the present study and is generally hard to solve
because of an intrinsic problem: Centaurus cluster samples are
typically dominated by Cen30 galaxies. In our Cen45 sample we have a
statistical number of $<$7 objects per morphological type which is far
away from being sufficient to get a good distance from e.g.~the
Tully-Fisher relation for spirals or the $D_n-\sigma$ relation for
ellipticals. An additional problem is the increasing background
contamination for samples taken from larger cluster areas which is
discussed in more details in Jerjen \& Dressler (1997b).  Apparently,
only with distances to individual galaxies it will be possible to
answer the distance question. In fact, Dressler (1993) determined
distances to six elliptical galaxies, 4 in Cen30 and 2 in Cen45
applying the surface brightness fluctuation method. The latter galaxies
(CCC$\,$130 and CCC$\,$134) have high probabilities of being Cen45
members: 0.76 and 0.66, respectively. He found no significant distance
difference between the six galaxies. This preliminary but important
result combined with our results on the physical nature of Cen30 and
Cen45 gives a consistent picture for the Centaurus cluster of a
dynamically young, unrelaxed system which is still in the process of
formation. Thereby the infalling galaxy group Cen45 represents a main
subcomponent which might be close to the cluster centre and thus
experience a high peculiar velocity.

In this paper, we have also investigated in detail on the very
different velocity distribution of early-type dwarfs as compared to
those of giant galaxies or Im\&BCD dwarfs. The early-type dwarfs are
highly concentrated towards the gravitational centre of the massive
Cen30 in a Gaussian-like distribution.  This stands in contrast to the
results for the Virgo cluster, where dE\&dS0s have a non-Gaussian
velocity distribution (Ferguson 1992).  This behaviour may give some
clue to the physical processes responsible for the formation and
evolution of dwarf ellipticals in clusters. Another signature for
ongoing evolution in the cluster core is represented by the detection
of a small bound population of early-type giant and dwarf galaxies
around the central dominant galaxy of Cen30. As a strong hint that a
few percent of the cluster galaxies are, in fact, bound either to
NGC$\,$4696 or to a peak in the cluster mass distribution, coinciding
with the position of NGC$\,$4696.

\begin{acknowledgements} 
It is a pleasure to thank Dr. B.~Binggeli for many comments and suggestions 
on the manuscript. We are grateful to the Swiss National Science Foundation
for financial support.
\end{acknowledgements}


\onecolumn
\footnotesize
\tabcolsep0.15cm

\clearpage

\begin{longtable}[l]{rllrllllccc}
\caption*{\raggedright{\bf Table\,5a. New redshifts for Centaurus
cluster members.} Coordinates (1950.0) are given in columns 2 and 3,
column 4 gives the Centaurus Cluster Catalogue (Jerjen \& Dressler
1996a) reference number. The galaxy morphological type (5), the total blue
magnitude (6) and effective surface brightness (7) are taken from this
same source, while the probability of being Cen45 cluster member (8)
has been derived in Section\,5. Please note that if prob$_{45}$=0.00 this 
automatically implies that the galaxy is a 100\% Cen30 member. Column 11 
contains the heliocentric radial velocity (cz) and its estimated 
uncertainty, which results from a weighted average of columns 9 
(cross-correlation redshift) and column 10 (emission line redshift).}\\
\hline
\multicolumn{1}{c}{Seq} & 
\multicolumn{1}{c}{RA} & 
\multicolumn{1}{c}{Dec} & 
\multicolumn{1}{c}{CCC} & 
\multicolumn{1}{c}{Type} &
\multicolumn{1}{c}{$B_{\rm T}$} & 
\multicolumn{1}{c}{{\it SB}$_{\rm eff}$}&
\multicolumn{1}{c}{prob$_{45}$}&
\multicolumn{1}{c}{V$_{\rm ccf}$} & 
\multicolumn{1}{c}{V$_{\rm emi}$} & 
\multicolumn{1}{c}{V}  \\
\multicolumn{1}{c}{(1)} & 
\multicolumn{1}{c}{(2)} & 
\multicolumn{1}{c}{(3)} & 
\multicolumn{1}{c}{(4)} & 
\multicolumn{1}{c}{(5)} & 
\multicolumn{1}{c}{(6)} &
\multicolumn{1}{c}{(7)} &
\multicolumn{1}{c}{(8)} &
\multicolumn{1}{c}{(9)} &
\multicolumn{1}{c}{(10)} &
\multicolumn{1}{c}{(11)} \\
\hline               
\endfirsthead

\caption*{(cont.)}\\
\hline                    
\multicolumn{1}{c}{Seq} & 
\multicolumn{1}{c}{RA} & 
\multicolumn{1}{c}{Dec} & 
\multicolumn{1}{c}{CCC} & 
\multicolumn{1}{c}{Type} &
\multicolumn{1}{c}{$B_{\rm T}$} & 
\multicolumn{1}{c}{{\it SB}$_{\rm eff}$}&
\multicolumn{1}{c}{prob$_{45}$}&
\multicolumn{1}{c}{V$_{\rm ccf}$} & 
\multicolumn{1}{c}{V$_{\rm emi}$} & 
\multicolumn{1}{c}{V}  \\
\multicolumn{1}{c}{(1)} & 
\multicolumn{1}{c}{(2)} & 
\multicolumn{1}{c}{(3)} & 
\multicolumn{1}{c}{(4)} & 
\multicolumn{1}{c}{(5)} & 
\multicolumn{1}{c}{(6)} &
\multicolumn{1}{c}{(7)} &
\multicolumn{1}{c}{(8)} &
\multicolumn{1}{c}{(9)} &
\multicolumn{1}{c}{(10)} &
\multicolumn{1}{c}{(11)} \\
\hline                    
\endhead
  1&12~44~44.68&$-$41~22~39.0&  1&dE,N  &17.09&22.7&0.00 & 3527$\pm$59&         & 3527$\pm$59\\
  3&12~44~47.86&$-$40~51~33.0&  6&Im    &16.65&23.2&0.00 &         & 3848$\pm$50& 3848$\pm$50\\
  5&12~44~59.61&$-$40~52~59.0& 11&dS0:N &17.75&23.3&0.00 & 2750$\pm$43&         & 2750$\pm$43\\
  7&12~45~03.75&$-$40~43~33.9& 13&dS0,N &16.83&22.5&0.00 & 3026$\pm$29&         & 3026$\pm$29\\
 13&12~45~22.85&$-$41~35~26.0& 32&Im    &16.98&23.5&0.00 & 2737$\pm$28&         & 2737$\pm$28\\
 15&12~45~29.58&$-$41~26~36.0& 38&dE,N  &18.10&23.9&0.00 & 3925$\pm$45&         & 3925$\pm$45\\
 16&12~45~30.24&$-$41~02~35.0& 41&Im    &18.75&23.9&0.00 &         & 4278$\pm$58& 4278$\pm$58\\
 22&12~45~50.31&$-$41~10~02.9& 58&dE    &17.94&24.1&0.00 & 3304$\pm$60&         & 3304$\pm$60\\
 23&12~45~53.96&$-$40~59~44.0& 61&dE,N  &17.43&23.8&0.00 & 2910$\pm$67&         & 2910$\pm$67\\
 28&12~46~14.60&$-$40~42~02.0& 73&SBm   &16.90&23.6&0.88 &         & 4823$\pm$50& 4823$\pm$50\\
 30&12~46~16.08&$-$40~59~14.0& 75&dE,N  &18.06&23.2&0.00 & 1958$\pm$71&         & 1958$\pm$71\\
 34&12~46~23.88&$-$40~55~14.0& 84&dS0   &17.73&23.7&0.00 & 3122$\pm$29&         & 3122$\pm$29\\
 36&12~46~28.05&$-$41~27~32.9& 88&dE,N  &18.06&    &0.00 & 2848$\pm$26&         & 2848$\pm$26\\
 38&12~46~36.81&$-$40~58~58.0& 91&dE    &19.40&24.8&0.00 & 3652$\pm$61&         & 3652$\pm$61\\
 42&12~46~44.21&$-$41~08~55.0& 97&dE,N  &18.70&23.7&0.00 & 2818$\pm$30&         & 2818$\pm$30\\
 47&12~46~53.03&$-$41~32~14.0&109&Im    &19.48&24.8&0.00 & 3545$\pm$60&         & 3545$\pm$60\\
 48&12~46~54.00&$-$41~05~38.9&111&dE,N  &16.94&23.8&0.00 & 2880$\pm$40&         & 2880$\pm$40\\
 49&12~46~56.04&$-$40~57~23.9&113&E     &17.33&22.6&0.00 & 3715$\pm$26&         & 3715$\pm$26\\
 52&12~47~06.75&$-$40~34~29.0&120&dS0   &17.00&23.3&0.00 & 3106$\pm$28&         & 3106$\pm$28\\
 53&12~47~08.06&$-$41~04~03.0&121&Im    &18.36&24.3&0.88 & 4739$\pm$70&         & 4739$\pm$70\\
 55&12~47~10.02&$-$41~07~43.9&123&dS0   &18.45&24.0&0.82 & 4661$\pm$69&         & 4661$\pm$69\\
 56&12~47~10.38&$-$40~59~16.9&125&dE,N  &17.14&23.8&0.00 & 2880$\pm$35&         & 2880$\pm$35\\
 64&12~47~34.73&$-$41~23~31.0&146&dE    &17.77&23.7&0.00 & 2675$\pm$53&         & 2675$\pm$53\\
 65&12~47~37.96&$-$41~01~28.9&150&dE,N  &18.23&24.5&0.13 & 4426$\pm$46&         & 4426$\pm$46\\
 71&12~48~12.00&$-$41~06~26.0&172&dE,N  &18.21&24.6&0.86 & 4844$\pm$45&         & 4844$\pm$45\\
 72&12~48~12.65&$-$41~13~22.9&173&Im    &18.30&24.1&0.00 & 2139$\pm$36&         & 2139$\pm$36\\
 82&12~48~51.43&$-$40~53~59.0&193&Sc    &17.51&23.4&0.76 & 4924$\pm$56&         & 4924$\pm$56\\
 83&12~48~53.30&$-$41~12~39.0&196&Im    &18.10&24.3&0.83 &         & 4879$\pm$58& 4879$\pm$58\\
 88&12~49~06.16&$-$41~26~10.0&209&dE    &18.32&24.2&0.00 & 3535$\pm$76&         & 3535$\pm$76\\
 91&12~49~14.16&$-$40~52~15.0&212&Im    &19.06&    &0.00 & 2085$\pm$39&         & 2085$\pm$39\\
 94&12~49~15.44&$-$41~08~49.9&215&dE    &19.50&24.6&0.00 & 3329$\pm$76&         & 3329$\pm$76\\
 96&12~49~18.24&$-$40~38~05.9&219&Im    &17.10&23.5&0.00 &         & 3284$\pm$71& 3284$\pm$71\\
103&12~49~34.13&$-$41~28~48.0&228&dE,N  &18.09&24.1&0.00 & 3212$\pm$62&         & 3212$\pm$62\\
104&12~49~35.37&$-$40~53~47.0&230&Im    &17.94&24.4&0.83 &         & 4881$\pm$45& 4881$\pm$45\\
106&12~49~48.10&$-$40~57~16.0&239&Im    &18.00&24.4&0.64 &         & 4971$\pm$71& 4971$\pm$71\\
112&12~51~15.00&$-$41~07~31.0&277&Sc    &16.54&22.1&0.17 & 4441$\pm$73&         & 4441$\pm$73\\
\hline
\end{longtable}

\clearpage

\begin{longtable}[l]{rllrllccc}
\caption*{\raggedright{\bf Table\,5b. New redshifts of background
galaxies.} Coordinates (1950.0) are given in columns 2 and 3,
column 4 gives the Centaurus Cluster Catalogue (Jerjen \& Dressler
1996a) reference number. The galaxy morphological type (5), the total blue
magnitude (6) and effective surface brightness (7) are taken from this
same source. Column 10 contains the
heliocentric radial velocity (cz) and its estimated uncertainty, which
results from a weighted average of columns 8 (cross-correlation
redshift) and column 9 (emission line redshift).}\\
\hline
\multicolumn{1}{c}{Seq} & 
\multicolumn{1}{c}{RA} & 
\multicolumn{1}{c}{Dec} & 
\multicolumn{1}{c}{CCC} & 
\multicolumn{1}{c}{$B_{\rm T}$} & 
\multicolumn{1}{c}{{\it SB}$_{\rm eff}$}&
\multicolumn{1}{c}{V$_{\rm ccf}$} & 
\multicolumn{1}{c}{V$_{\rm emi}$} & 
\multicolumn{1}{c}{V}  \\
\multicolumn{1}{c}{(1)} & 
\multicolumn{1}{c}{(2)} & 
\multicolumn{1}{c}{(3)} & 
\multicolumn{1}{c}{(4)} & 
\multicolumn{1}{c}{(5)} & 
\multicolumn{1}{c}{(6)} &
\multicolumn{1}{c}{(7)} &
\multicolumn{1}{c}{(8)} &
\multicolumn{1}{c}{(9)} \\
\hline          
\endfirsthead

 6&12~45~00.98&$-$40~56~44.0& 12&18.82&24.4&38469$\pm$66&         &38469$\pm$66\\
14&12~45~26.04&$-$40~16~07.9& 33&17.49&23.1&14702$\pm$81&14706$\pm$58&14705$\pm$47\\
18&12~45~36.96&$-$41~25~31.0& 44&19.76&24.3&         &21123$\pm$50&21123$\pm$50\\
21&12~45~49.10&$-$41~27~29.9& 56&18.90&24.8&13792$\pm$84&         &13792$\pm$84\\
29&12~46~15.02&$-$41~09~21.0& 74&18.76&23.7&16514$\pm$64&         &16514$\pm$64\\
39&12~46~38.54&$-$40~21~50.0& 92&17.70&23.5&         &10746$\pm$58&10746$\pm$58\\
57&12~47~14.58&$-$41~27~06.0&127&17.85&23.6&16362$\pm$45&         &16362$\pm$45\\
58&12~47~15.19&$-$41~10~30.0&128&19.28&24.2&46615$\pm$85&         &46615$\pm$85\\
63&12~47~31.28&$-$41~01~54.9&142&18.40&24.0&20843$\pm$56&         &20843$\pm$56\\
74&12~48~27.17&$-$40~32~34.9&179&17.27&22.6&14863$\pm$61&14878$\pm$58&14871$\pm$42\\
76&12~48~31.57&$-$41~00~13.0&182&18.01&23.3&         &10679$\pm$38&10679$\pm$38\\
77&12~48~41.55&$-$40~27~12.9&186&17.98&22.6&27387$\pm$91&27371$\pm$58&27376$\pm$49\\
92&12~49~14.18&$-$40~18~54.0&213&17.23&23.5&15916$\pm$44&         &15916$\pm$44\\
97&12~49~18.44&$-$41~22~21.0&220&17.93&23.1&         &19439$\pm$50&19439$\pm$50\\
\hline
\end{longtable}

\clearpage 

\begin{longtable}[l]{rllrllllccc}
\caption*{\raggedright{\bf Table\,5c. Reobserved redshifts for
Centaurus cluster members.} Coordinates (1950.0) are given in columns 2
and 3, column 4 gives the Centaurus Cluster Catalogue (Jerjen \&
Dressler 1996a) reference number. The galaxy morphological type (5),
the total blue magnitude (6) and effective surface brightness (7) are
taken from this same source, while the probability of being Cen45
cluster member (8) has been derived in Section\,5. Please note that if
prob$_{45}$=0.00 this automatically implies that the galaxy is a 100\%
Cen30 member. Column 11 contains the heliocentric radial velocity (cz)
and its estimated uncertainty, which results from a weighted average of
columns 9 (cross-correlation redshift) and column 10 (emission line
redshift).}\\ \hline 
\multicolumn{1}{c}{Seq} & \multicolumn{1}{c}{RA} &
\multicolumn{1}{c}{Dec} & \multicolumn{1}{c}{CCC} &
\multicolumn{1}{c}{Type} & \multicolumn{1}{c}{$B_{\rm T}$} &
\multicolumn{1}{c}{{\it SB}$_{\rm eff}$}&
\multicolumn{1}{c}{prob$_{45}$}& \multicolumn{1}{c}{V$_{\rm ccf}$} &
\multicolumn{1}{c}{V$_{\rm emi}$} & \multicolumn{1}{c}{V} \\
\multicolumn{1}{c}{(1)} & \multicolumn{1}{c}{(2)} &
\multicolumn{1}{c}{(3)} & \multicolumn{1}{c}{(4)} &
\multicolumn{1}{c}{(5)} & \multicolumn{1}{c}{(6)} &
\multicolumn{1}{c}{(7)} & \multicolumn{1}{c}{(8)} &
\multicolumn{1}{c}{(9)} & \multicolumn{1}{c}{(10)} &
\multicolumn{1}{c}{(11)} \\ \hline \endfirsthead

\caption*{(cont.)}\\
\hline               
\multicolumn{1}{c}{Seq} & 
\multicolumn{1}{c}{RA} & 
\multicolumn{1}{c}{Dec} & 
\multicolumn{1}{c}{CCC} & 
\multicolumn{1}{c}{Type} &
\multicolumn{1}{c}{$B_{\rm T}$} & 
\multicolumn{1}{c}{{\it SB}$_{\rm eff}$}&
\multicolumn{1}{c}{prob$_{45}$}&
\multicolumn{1}{c}{V$_{\rm ccf}$} & 
\multicolumn{1}{c}{V$_{\rm emi}$} & 
\multicolumn{1}{c}{V}  \\
\multicolumn{1}{c}{(1)} & 
\multicolumn{1}{c}{(2)} & 
\multicolumn{1}{c}{(3)} & 
\multicolumn{1}{c}{(4)} & 
\multicolumn{1}{c}{(5)} & 
\multicolumn{1}{c}{(6)} &
\multicolumn{1}{c}{(7)} &
\multicolumn{1}{c}{(8)} &
\multicolumn{1}{c}{(9)} &
\multicolumn{1}{c}{(10)} &
\multicolumn{1}{c}{(11)} \\
\hline                 
\endhead
 
  2&12~44~46.32&$-$40~59~22.0&  4&S0    &16.93&22.1&0.00& 2704$\pm$61& 2714$\pm$58& 2709$\pm$42\\
  4&12~44~56.98&$-$41~15~21.9&  8&SB0   &14.03&21.5&0.00& 3531$\pm$25&         & 3531$\pm$25\\
  8&12~45~10.56&$-$40~38~53.0& 17&S0    &15.95&22.4&0.00& 3874$\pm$70&         & 3874$\pm$70\\
  9&12~45~14.22&$-$40~56~49.2& 19&d:S0  &15.82&22.1&0.00& 4059$\pm$19&         & 4059$\pm$19\\
 10&12~45~16.54&$-$41~01~57.5& 22&dE    &17.94&23.5&0.00& 2433$\pm$33&         & 2433$\pm$33\\
 11&12~45~16.76&$-$40~19~19.0& 24&SBd   &15.21&22.5&0.00&         & 4017$\pm$58& 4017$\pm$58\\
 12&12~45~19.01&$-$40~56~29.9& 26&SBc   &16.06&22.4&0.00& 1414$\pm$77&         & 1414$\pm$77\\
 17&12~45~35.56&$-$41~26~28.9& 43&S0    &14.05&20.6&0.00& 2647$\pm$25&         & 2647$\pm$25\\
 19&12~45~37.24&$-$40~51~03.2& 45&E     &14.93&20.8&0.00& 3681$\pm$49&         & 3681$\pm$49\\
 20&12~45~45.33&$-$41~02~02.8& 54&E     &15.90&20.3&0.00& 2958$\pm$44&         & 2958$\pm$44\\
 24&12~45~57.41&$-$41~22~15.9& 62&S0    &15.37&21.8&0.00& 2193$\pm$19&         & 2193$\pm$19\\
 25&12~46~03.51&$-$41~02~18.9& 65&E/S0  &11.32&22.6&0.00& 2985$\pm$31&         & 2985$\pm$31\\
 26&12~46~08.09&$-$41~02~48.0& 70&E     &15.64&18.3&0.00& 2317$\pm$20&         & 2317$\pm$20\\
 27&12~46~09.66&$-$40~18~33.0& 71&S0    &14.76&21.3&0.00& 3541$\pm$25&         & 3541$\pm$25\\
 31&12~46~17.22&$-$41~07~10.7& 76&dS0   &16.41&21.8&0.00& 5298$\pm$28&         & 5298$\pm$28\\
 32&12~46~18.34&$-$41~03~59.9& 77&Sc    &15.99&23.2&0.00& 3513$\pm$34&         & 3513$\pm$34\\
 33&12~46~20.34&$-$41~33~35.9& 80&S0    &14.78&21.8&0.13& 4422$\pm$23&         & 4422$\pm$23\\
 35&12~46~26.03&$-$41~16~21.5& 85&SBa   &15.07&22.5&0.00& 2839$\pm$20&         & 2839$\pm$20\\
 37&12~46~32.63&$-$41~03~48.0& 89&E     &15.77&20.0&0.00& 3104$\pm$19&         & 3104$\pm$19\\
 40&12~46~39.53&$-$41~09~25.0& 94&dS0,N &16.67&23.2&0.00& 4076$\pm$29&         & 4076$\pm$29\\
 41&12~46~40.18&$-$41~13~03.0& 95&SB0   &14.56&21.7&0.00& 4222$\pm$21&         & 4222$\pm$21\\
 43&12~46~46.81&$-$41~37~43.0&102&S0    &14.65&22.4&0.00& 3347$\pm$23&         & 3347$\pm$23\\
 44&12~46~48.55&$-$40~46~59.0&103&S0    &14.41&20.4&0.00& 1990$\pm$24&         & 1990$\pm$24\\
 45&12~46~51.82&$-$41~06~57.1&106&SB0/a &14.62&21.7&0.00& 3677$\pm$19&         & 3677$\pm$19\\
 46&12~46~52.56&$-$40~48~58.7&108&E     &16.11&19.9&0.00& 3263$\pm$36&         & 3263$\pm$36\\
 50&12~46~56.15&$-$40~24~51.9&114&SBb   &15.78&23.0&0.00& 2242$\pm$24&         & 2242$\pm$24\\
 51&12~47~05.57&$-$40~57~14.0&119&E     &14.64&20.3&0.00& 2101$\pm$23&         & 2101$\pm$23\\
 54&12~47~08.12&$-$41~00~26.3&122&S0/a  &14.18&21.6&0.00& 3845$\pm$23&         & 3845$\pm$23\\
 59&12~47~17.90&$-$41~06~35.9&130&E     &12.43&22.6&0.76& 4624$\pm$25&         & 4624$\pm$25\\
 60&12~47~21.59&$-$41~07~32.9&134&E     &15.04&20.6&0.66& 4966$\pm$31&         & 4966$\pm$31\\
 61&12~47~25.40&$-$40~56~57.2&135&E     &15.35&20.6&0.00& 2908$\pm$26&         & 2908$\pm$26\\
 62&12~47~26.12&$-$41~14~36.0&137&S0    &14.03&20.4&0.00& 2384$\pm$24&         & 2384$\pm$24\\
 66&12~47~38.33&$-$40~34~26.9&151&Sa    &14.27&22.2&0.79& 4640$\pm$20&         & 4640$\pm$20\\
 67&12~47~48.05&$-$41~11~58.0&158&S0    &14.77&21.3&0.01& 5249$\pm$24&         & 5249$\pm$24\\
 68&12~48~00.14&$-$41~37~20.9&165&S0    &14.87&20.7&0.89& 4758$\pm$25&         & 4758$\pm$25\\
 69&12~48~11.13&$-$41~07~28.9&170&S0    &15.24&20.4&0.25& 4466$\pm$19&         & 4466$\pm$19\\
 70&12~48~11.30&$-$40~36~53.0&171&BCD   &16.76&22.3&0.60& 4562$\pm$22&         & 4562$\pm$22\\
 73&12~48~14.65&$-$41~27~01.9&175&E     &14.65&21.7&0.00& 2469$\pm$23&         & 2469$\pm$23\\
 75&12~48~31.55&$-$40~32~50.0&181&SB0   &14.76&21.8&0.00& 3391$\pm$23&         & 3391$\pm$23\\
 78&12~48~42.69&$-$41~34~20.9&187&d:S0  &15.99&22.3&0.00& 1826$\pm$19&         & 1826$\pm$19\\
 79&12~48~46.40&$-$40~57~21.5&189&Sc    &15.02&21.3&0.89& 4771$\pm$24&         & 4771$\pm$24\\
 80&12~48~49.76&$-$41~13~14.4&190&dS0,N &16.11&21.7&0.00& 4048$\pm$23&         & 4048$\pm$23\\
 81&12~48~50.78&$-$41~01~54.7&191&S0    &15.32&21.1&0.00& 3467$\pm$22&         & 3467$\pm$22\\
 84&12~48~59.52&$-$40~59~42.7&203&dS0   &16.72&21.7&0.08& 4400$\pm$20&         & 4400$\pm$20\\
 85&12~49~01.57&$-$40~43~19.9&205&S0    &15.95&22.1&0.00& 2563$\pm$47&         & 2563$\pm$47\\
 86&12~49~04.27&$-$40~54~52.8&207&E     &16.01&21.0&0.87& 4712$\pm$18&         & 4712$\pm$18\\
 87&12~49~04.70&$-$41~09~38.3&208&dE,N  &17.44&23.3&0.00& 3624$\pm$22&         & 3624$\pm$22\\
 89&12~49~09.74&$-$41~16~02.8&210&SB0   &15.16&21.7&0.00& 3675$\pm$19&         & 3675$\pm$19\\
 90&12~49~13.97&$-$40~52~32.2&211&S0    &14.09&22.1&0.00& 2025$\pm$26&         & 2025$\pm$26\\
 93&12~49~14.56&$-$41~00~28.0&214&Sc    &14.73&22.6&0.89& 4806$\pm$32& 4757$\pm$50& 4792$\pm$27\\
 95&12~49~16.48&$-$41~11~17.6&217&SB0   &14.04&21.1&0.00& 3950$\pm$22&         & 3950$\pm$22\\
 98&12~49~26.21&$-$41~04~02.8&222&Sc    &14.74&23.2&0.17& 5107$\pm$31&         & 5107$\pm$31\\
 99&12~49~28.70&$-$41~12~19.0&224&dS0   &16.49&23.5&0.00& 3037$\pm$26&         & 3037$\pm$26\\
100&12~49~29.06&$-$40~59~15.5&225&S0    &15.71&22.5&0.00& 3459$\pm$23&         & 3459$\pm$23\\
101&12~49~29.31&$-$41~07~08.6&226&S0/a  &14.08&21.7&0.00& 2968$\pm$23&         & 2968$\pm$23\\
102&12~49~33.02&$-$40~47~22.0&227&SBa   &13.60&22.2&0.00& 3389$\pm$26&         & 3389$\pm$26\\
105&12~49~39.54&$-$40~26~13.0&236&SB0   &14.32&22.0&0.18& 4445$\pm$21&         & 4445$\pm$21\\
107&12~49~54.27&$-$40~57~31.0&242&S0    &15.46&19.3&0.85& 4858$\pm$23&         & 4858$\pm$23\\
108&12~50~18.65&$-$40~33~13.9&255&S0    &15.92&20.6&0.87& 4831$\pm$22&         & 4831$\pm$22\\
109&12~50~27.27&$-$40~52~10.6&260&dE,N  &17.59&23.9&0.00& 2766$\pm$43&         & 2766$\pm$43\\
110&12~50~39.31&$-$40~55~59.9&266&E     &13.08&21.4&0.01& 4314$\pm$23&         & 4314$\pm$23\\
111&12~50~55.53&$-$40~45~32.0&272&Im/BCD&16.51&23.0&0.00& 4215$\pm$48&         & 4215$\pm$48\\
113&12~51~19.19&$-$40~37~47.9&279&S0    &15.67&20.7&0.00& 3024$\pm$17&         & 3024$\pm$17\\
114&12~51~41.30&$-$40~53~12.0&284&Sd    &15.86&21.7&0.02& 4367$\pm$41& 4319$\pm$58& 4351$\pm$34\\
115&12~52~10.47&$-$40~42~48.9&290&E     &16.31&21.7&0.02& 4344$\pm$19&         & 4344$\pm$19\\
\hline
\end{longtable}


\begin{thebibliography}{} 

\bibitem{}
Binggeli, B., 1993, Habilitationsschrift, University of Basel

\bibitem{}
Binggeli, B., Sandage, A., Tammann, G.A., 1987, AJ, 94, 251 
 
\bibitem{}
Binggeli, B., Tarenghi, M., Sandage, A., 1990, A\&A, 228, 42
 
\bibitem{}
Binggeli, B., Popescu, C.C., Tammann, G.A., 1993, A\&AS, 98, 275
 
\bibitem{}
Blakeslee, J.P., Tonry, J.L., 1992, AJ, 103, 1457

\bibitem{}
Carollo, C.M., Danziger, I.J., Buson, L., 1993, MNRAS, 265, 553

\bibitem{}
Cuby, J.-G., Mignoli, M., 1994, in ``Instrumentation in Astronomy
VIII'' SPIE Conf. Ser.\,2198, pag. 98

\bibitem{}
Dickens, R.J., Currie, M.J., Lucey, J.R., 1986, MNRAS, 220, 679 (DCL)
 
\bibitem{} 
Dressler, A., 1980, ApJ, 236, 351
 
\bibitem{}
Dressler, A., 1993, in "Cosmic Velocity Fields", Proceedings of the
9th IAP Astrophysics Meeting, eds. F.R. Bouchet and M. Lachi\`eze-Rey,
Editions Frontieres, Gif-sur-Yvette, p.\,9
 
\bibitem{}
Dressler A., Faber S.M., Burstein D., Davies R.L., Lynden-Bell D., 
Terlevich R.J., Wegner G., 1987, ApJ, 313, L37

\bibitem{}
Faber S.M., Wegner G., Burstein D., Davies R.L., Dressler A., 
Lynden-Bell D., Terlevich R.J., ApJS, 1989, 69, 763

\bibitem{}
Felenbok, P. \et, 1997, Experimental Astronomy, 7, 65

\bibitem{}
Ferguson, H.C., 1992, MNRAS, 255, 389

\bibitem{}
Ferguson, H.C., Sandage, A., 1988, AJ, 96, 1520 

\bibitem{}
Ferguson, H.C., Sandage, A., 1991, AJ, 101, 765 
 
\bibitem{}
Gebhardt, K., Beers, T. C., 1991, ApJ, 383, 72

\bibitem{}
Girardi, M., Biviano, A., Giuricin, G., Mardirossian, F., Mezzetti, M.,
1993, ApJ, 404, 38

\bibitem{}
Held, E.V., Mould, J.R., 1994, AJ, 107, 1307
 
\bibitem{}
Horne K., 1986, PASP 98, 609

\bibitem{} 
Jerjen, H., 1995, PhD thesis, Astronomical Institute of the University 
of Basel, Switzerland
 
\bibitem{}
Jerjen, H., 1997, in preparation   

\bibitem{}
Jerjen, H., Dressler, A., 1997a, A\&AS, in press   
 
\bibitem{}
Jerjen, H., Dressler, A., 1997b, in preparation   

\bibitem{}
Jerjen, H., Tammann, G.A., 1997, A\&A, 321, 713

\bibitem{}
Jerjen, H., Freeman, K.C., Binggeli, B., 1997, in preparation
 
\bibitem{}
Kraan-Korteweg R.C., Woudt P.A., Cayatte V., Fairall A.P., Balkowski
C., Henning P.A., 1996, Nature, 379, 519 

\bibitem{}
Lissandrini C., Cristiani S., La Franca F., 1994, PASP, 106, 1157L

\bibitem{}
Lucey, J.R., Carter, D., 1988, MNRAS, 235, 1177
 
\bibitem{}
Lucey, J.R., Currie, M.J., Dickens R.J., 1986a, MNRAS, 221, 453 (LCD)
 
\bibitem{}
Lucey, J.R., Currie, M.J., Dickens R.J., 1986b, MNRAS, 222, 427
 
\bibitem{}
Lynden-Bell, D., Faber, S.M., Burstein, D., Davies, R.L., Dressler, A., Wegner, G., 1988, ApJ, 326, 19
 
\bibitem{}
Merrifield, M.R., Kent, S.M., 1991, AJ, 101, 783

\bibitem{}
Sandage, A., Binggeli, B., Tammann, G.A., 1985, AJ, 90, 1759

\bibitem{}
Stein, P.,  1996, A\&AS, 116, 203

\bibitem{}
Stein, P.,  1997, A\&A, 317, 670

\bibitem{}
Tonry, J., Davis, M., 1979, AJ, 84, 1511 

\bibitem{}
Vader, J.P., Sandage, A., 1991, ApJ, 379, L1

\end{thebibliography}
\end{document}